\documentclass[english]{paper} 
\usepackage[top=1.0in,bottom=1.0in,left=0.6in,right=0.6in]{geometry} 
\usepackage[unicode=true,bookmarks=true,bookmarksnumbered=false,bookmarksopen=true,bookmarksopenlevel=1,breaklinks=false,pdfborder={0 0 1},backref=section,colorlinks=false]{hyperref} 
\usepackage{amsmath} 
\usepackage{enumerate} 
\usepackage[numbers, sort&compress]{natbib} 
\usepackage{verbatim} 
\usepackage{graphicx} 
\usepackage{subcaption}
\newcommand{\hidsec}[1]{\par} 

\begin{document}
\title{A United Framework for both Formal, Natural and Social Science} 
\author{SHEN Fan\thanks{e-mail:shenfan19@gmail.com}} 
\date{}
\maketitle

\hidsec{abstract}

\begin{abstract} 
    With the expansion of scientific research, the number of scientific research is increasing. A new urgent problem is raised that how to keep these researches in a proper way.
    Therefore, knowledge mapping methods come into being, providing a lot of management and application functions. 
    However, it is still a problem to fully understand the knowledge map, especially in the field of sociology, which is always seperated with natural science. 
    To answer this question, a three-dimensional knowledge map is proposed in this paper, with time, space and number based on category and numericity. It is able to conclude most of scientific problems related to numericity interdisciplinary.
    Compared with the traditional way, this map is normative, and puts forward the general production criteria of labeling and digitization. It is also intuitive and readable, on which nature, society and formal science are expressed in the same view. The scientific methodologies are summarized, so that the methods with similar logic between different disciplines can be used for reference in development.
    Some social subjects are expressed more vividly than traditional text-based expressions, and are compatible with the natural science system.
    Mathematics also show its importance on the map as formal Science, indicating that it is the key to the development of science.
    This is not only a preliminary model of a comprehensive scientific worldview, but also a preliminary framework for the connection and cooperation of various disciplines in the future.

\end{abstract}

\section{Introduction}
\hidsec{why category}
    With the development of society, the accumulation of scientific and technological knowledge is increasing, and there are more and more interdisciplinary subjects\cite{BornerMakingSenseMankind2007,PorterWheredoesnanotechnology2009,Klavansconsensusmapscience2009}.
    The cost of knowledge classification, learning, retrieval and application is also getting higher.
    Scientists have been trying to find a way to classify knowledge\cite{PrasetyaStudyExtendedScratchBuild2020,Leydesdorffglobalmapscience2009}, but there is no systematic objective method at present.
\hidsec{current category}
    A long time ago, under the simple category theory, Ionian philosophy consider there are 4 basic elements of matter, while ancient China divided them into 5 elements.
    After that, philosophy of science rose, Aristotle, Kant\cite{kantCritiquePureReason1908}, Hegel\cite{hegelWissenschaftLogikObjective1816} and more scientists made contribution in their own ways.
    In order to provide a reliable theoretical foundation for modern science, philosophers seek a comprehensive philosophy of science\cite{popperLogicScientificDiscovery2005,matthewsMarioBungeSystematic2012}.
    At the same time, social sciences are moving towards quantification\cite{Saltellisociologyquantificationethics2020}, religious philosophy is also seeking new deconstruction, integration and progress\cite{BidwellReligiousDiversityPublic2015,GunaratneGoEastyoung2013}, although they are too complex to construct recognized frameworks.
\hidsec{modern category}
    With the advent of information explosion era, the task of scientific knowledge map has gradually changed from philosophical discussion to numerical summarization\cite{GOFFEYconceptmapbasedknowledge2002,Pardosuniversitymapcourse2020}.
    Knowledge map is becoming popular gradually, and its drawing is more completed by information science, knowledge engineering and informatics\cite{huangKnowledgeMapVisualization,CoboSciencemappingsoftware2011}.
    The focus of scientific attention began to fall on each specific discipline, and the cognitive differences between them became bigger\cite{ColeWhysociologydoesn1994}, which reduced the level of overall thinking\cite{NicholasMaxwellNewTaskPhilosophy2019}.
\hidsec{this method}
    This paper will try to build a framework from the engineering perspective with the simplest mathematical and categorical thinking, so as to understand the nature and mission of science conveniently.
    At the same time, the numerical framework can be managed by computer, probably will also be analyzed by machine learning in the future.
    In this way, scientific research can be carried out more clearly and effectively under the guidance of logic and philosophical methodology.

\section{Definition of Coordinate System}
\hidsec{tagging}
    In this paper, the three basic dimensions of time, space and quantity are adopted, and the logarithmic coordinate axis is used to construct the three-dimensional coordinate system considering the scale differences among various disciplines.
    Therefore, an object can be located on a coordinate system as
    \begin{equation} \label{eq:tsq_axis}
    X = \left[ \begin{array}{ccc}
    T & S & Q \end{array} \right]^{T}
    = \lg \left[ \begin{array}{ccc}
    t & s & q \end{array}\right]^{T}
    \end{equation}

    In Eq \eqref{eq:tsq_axis}, $t$ is a measured time in seconds, $s$ is a measured length in meters and $q$ is a quantity, number or population. The unit is constant $1$.
    In this paper, this system is called as Time-Space-Quantity Coordinate System (TSQ-CS). 
    Each discipline is located to the coordinate axis with the numerical attribute of its research object. For example, the time quantity can be selected as the overall life or cycle period, and the space quantity can be selected as its own diameter, orbit diameter, movement distance, etc.
    In particular, in order to distinguish the composition of each other, taking the human view as observer, it can be note as
    \begin{equation}
        \begin{cases}
        q = n, & \text{when 1 object is composed with n observers, like company or country}\\
        q = 1/n, & \text{when n objects composed to 1 observer, like cells or atoms}
        \end{cases}
        \label{eq:def_q}
    \end{equation}
    Therefore, it could be obtained from \eqref{eq:def_q} that $Q^- = -\lg q$ and $Q^+ = \lg q$. Hence, the Q-axis is properly defined.

    Due to the limitations of the author's knowledge scope and working time, this paper only selects some typical disciplines and puts them into the coordinate system as examples, as shown in table \ref{tab:all_values}.
\hidsec{table TSQ: newer than sheet, need to match}
    \begin{table}[htbp]
    \caption{Coordinate values of disciplines/objects in TSQ-CS}
    \label{tab:all_values}
    \centering
    \begin{tabular}{llllll}
    \hline
    Discipline & Sample & $t(s)$ & $s(m)$ & $q(1)$ & Reference\\
    \hline
    Classical mechanics    & Human         & $1.0$ & $1.0$ & $1.0$ & A cycle of human  \\
    Galactic astronomy & Galaxy & $7.1\times 10^{15}$ & $1.0\times 10^{21}$ & $2.0\times 10^{12}$ & A galactic year \\
    Planetary science  & Planet earth  & $8.6\times 10^4$ & $1.3\times 10^7$ & 1.0 & Rotation of Earth  \\
    Stellar astronomy   & Star sun   & $2.1\times 10^6$ & $1.4\times 10^9$ & 1.0 & Rotation of Sun    \\
    Biology & Macro molecular & $1.2\times 10^1$ & $1.0\times 10^{-9}$ & $1.0\times 10^{-27}$ &Estimated\\
    Chemistry & Atom  & $4.0\times 10^{-7}$ & $1.1\times 10^{-10}$ & $1.0\times 10^{-28}$ & Bohr radius \\
    Chemistry & Molecular & $1.2\times 10^1$ & $7.4\times 10^{-11}$ & $1.0\times 10^{-27}$ & Atom/3 estimated \\
    Engineering   & Machinery big   & $3.2\times 10^8$ & $3.0\times 10^2$ & $1.0\times 10^2$ & 10 years estimated \\
    Engineering  & Machinery small  & $1.6\times 10^7$ & $1.0\times 10^{-3}$ & 1.0 & 0.5 year estimated \\
    Medicine   & Organ/Cell  & $2.6\times 10^5$ & $2.0\times 10^{-5}$ & $1.0\times 10^{-14}$ & Villi of intestine\\
    Optics  & Light & $1.0\times 10^{13}$ & $3.0\times 10^{21}$ & 1.0 & from $s/t = c$ \\
    Optics & Photon & $1.3\times 10^{-15}$ & $2.8\times 10^{-15}$ & $1.0\times 10^{-28}$ & Violet light \\
    Quantum mechanics  & Electron     & $4.5\times 10^{-7}$ & $2.8\times 10^{-15}$ & $1.0\times 10^{-30}$ & Wikipedia\\
    Quantum mechanics  & Virtual quantum & $1.8\times 10^{-2}$ & $1.1\times 10^{-10}$ & $1.0\times 10^{-30}$ & Electron orbit \\
    Macroeconomics & Country  & $3.2\times 10^7$ & $5.6\times 10^5$ & $6.8\times 10^7$ & A country(UK)\\
    Microeconomics  & Person        & $3.2\times 10^7$ & $1.5\times 10^1$ & $1.0$ & A people \\
    Politics   & Capitalism  & $2.6\times 10^6$ & $1.0\times 10^2$ & $2.0\times 10^2$ & A company\\
    Politics               & Libertarianism      & $2.6\times 10^6$ & 5.0 & 1.0 & A people  \\
    Politics   & Socialism  & $2.6\times 10^6$ & $5.6\times 10^5$ & $6.8\times 10^7$ & 3 month   \\
    Philosophy of religion & Buddhism sample& $1.0$ & $1.0$ & $2.1\times 10^{-5}$ & Buddhism x60n1115 \\
    Philosophy of religion & Taoism sample  & $2.2\times 10^9$ & 1.0 & 1.0 & Tao Te Ching   \\
    Sociology (Marxism)&Country& $8.6\times 10^4$& $5.6\times 10^5$ & $6.8\times 10^7$ & A country(UK)\\
    Sociology (Marxism)&Company& $8.6\times 10^4$& $1.0\times 10^2$ & $2.0\times 10^2$ & A company  \\
    Sociology (Marxism) & People          & $8.6\times 10^4$ & $5.0$ & $1.0$ & A people  \\
    \hline
    \end{tabular}
    \end{table}
\hidsec{figure TSQ}
    \begin{figure}[htbp]
    \centering
    \includegraphics[height=0.3\linewidth]{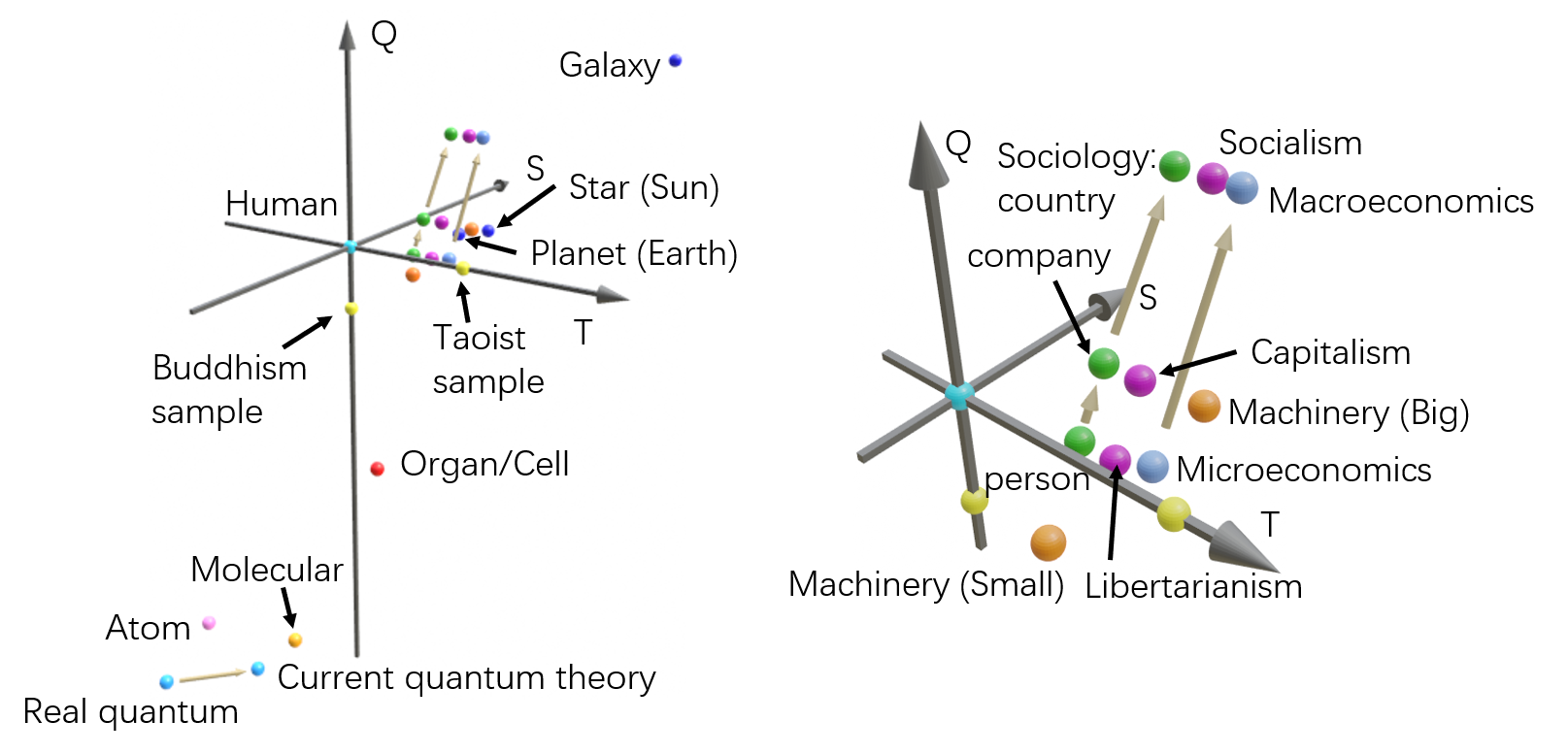}
    \caption{Natural, Formal and Social Sciences in TSQ-CS}
    \label{fig:map_all}
    \end{figure}
\hidsec{for natural science}
    The final form of knowledge map is shown in the figure \ref{fig:map_all}. It can be seen that the coordinate system represents a number of disciplines of natural science, including basic sciences such as physics and biology, as well as applied sciences such as engineering and medicine, which mostly located as $Q<0$.
    By the same time, due to the establishment of innovative quantitative axis, social sciences also get objective expression in numerical level, most of them are located as $Q>0$.

    Statistics in formal science is represented by arrows. As a typical mathematical operation, it exists in various disciplines of quantum mechanics and sociology in physics.
    Obviously, the coordinate system provides a new possibility for scientific induction.

\section{Categorization of Science}
\hidsec{brief}
    This section will discuss each discipline separately, which is divided into three parts: natural, formal and social science.
\subsection{Natural Science}
    Most disciplines of natural science can be positioned in TSQ-CS as figure \ref{fig:map_all}.
    Their general similarities are $Q < 0$, and $T,S \rightarrow 0$.
    Generally speaking, the closer to the origin, the more mature and clear the theory is.
\subsubsection{Engineering}
    Engineering machinery is the most mature theory and the most successful application subject. Compared with other disciplines, it is closest to the plane $Q=0$ and not distanced from the observer in scale.
    Therefore, a simple conclusion is that the closer to the $Q=0$ plane, and the smaller the distance to the observer's origin, the simpler the subject will be.
\subsubsection{Chemistry, Biology and Medic}
    Compared with engineering science, the development of medicine, biology and chemistry to the micro part is more difficult. 
    To a certain extent, medicine can be regarded as an applied discipline of biology\cite{KhalilSyntheticbiologyapplications2010}, both of which are extremely difficult due to the large number of irregular cells in research and application.  
    Compared with the two, chemistry is more tiny. 
    This undoubtedly proves the importance of Q-axis in the graph, that is, the number of research objects is an important indicator for the classification of disciplines. 
\subsubsection{Physics and Astronomy}
    The scales of modern physics from quantum mechanics to astronomy can be expressed in TSQ-CS.
    With the invention of tools, the scope of physics is expanding both in micro and macro\cite{FertMagneticskyrmionsadvances2017,MizunoJCMTopensnew2021}.

    The current quantum formula is given in the form of statistical probability\cite{HallQuantumtheorymathematicians2013}, which is different from most engineering and astronomical formulas.
    According to the thought of TSQ-CS, this paper can double-locate quantum mechanics, one of which is located in the real quantum space-time scale that can not be observed accurately by human beings, and the other is located in the statistical space-time scale that can be observed accurately by human beings, which is a virtual point.
    Therefore, it can be assumed that the current quantum mechanics observation does not reach the real scale of quantum time and space, but only to reach the scale that can be studied by human beings, so it is a statistical formula.
    Therefore, Schrodinger's cat paradox can be explained intuitively, that is, the half dead cat only sits on the virtual point, and the dead or alive of the cat in the real point can be obtained from the observation of an exact time.
\subsection{Formal Science}
    TSQ-CS can also calibrate formal science.
    As can be seen from the figure \ref{fig:sci_definition}, changes in time, space and quantity are involved in natural science research or engineering applications.
    This change is reflected in the movement of each coordinate axis direction, and the moving tool is mathematics.
    In the view of this paper, formal science themselves do not always have clear research object in the world, but are essential tools or rules of science, which run through all aspects.
    They reorganize the information of the research object and transmit it to the observer, such as statistics, which reorganizes the economic data of the group into a single concluded data for the observer.

    In particular, on the Q-axis, due to the discrete characteristics of the research object, the downward differential operation can not be carried out, so it can only be counted upward.
    This may be one of the reasons why the social science had not created close connection with natural sciences, because it is unable to do reliable transform from high-level to lower.
    At present, there is no specific solution to this problem in mathematics. The hot machine learning optimization nowadays may be one of the solution for it.
\subsection{Social Science}
    Social sciences include sociology, economics, politics and so on. They are often difficult to be compatible with the natural science, because most of their research objects are groups rather than individuals.
    The Q-axis of TSQ-CS in this paper can express the number of individuals\cite{ESPELANDSociologyQuantification2008,WeidlichConceptsModelsQuantitative2012}.
    For the convenience of observation in TSQ-CS, this paper defines its T-value artificially, i.e. 1 year in economics, 1 month in politics and 1 day in sociology.
\subsubsection{Economy}
    Economics is a subject that is relatively close to data in social science, so that several typical branches can be demarcated according to the attributes of their research objects.
    As the research object of macroeconomics is a country and society, and the research object of microeconomics is individuals and companies, the calibration of these 2 disciplines could be shown in the enlarged part of figure \ref{fig:map_all}.
    Obviously, there is no distinction between primary and secondary, and statistical methods can become a tool for connecting.
    This idea of local and global research is very similar to physics and industry, which is in line with the current general scientific logic, such as finite element method.
    However, there is no mature industrial method from macro to micro at present,, only through probability theory and optimization ideas such as game theory and dynamic programming to solve the ideal value, which needs further development of mathematics mentioned in the last section.
\subsubsection{Politics}
    For politics, the ideology of comparative politics is selected to discuss in this paper.
    liberalism, capitalism and socialism are most typical ideologies, which respectively attach importance to the individual, capital and the whole society, and each has its own advantages.
    As shown in the figure \ref{fig:map_all}, these respective focuses of ideologies can be clearly demarcated, and the differences are the focuses of some conflicts and wars in the world.
    Scientists may be able to help find a balanced and stable solution from the perspective of optimization theory, so as to avoid the political form itself becoming the fuse of interest disputes,
    even make the national policy more reasonable and effective.
\subsubsection{Sociology}
    The research field of sociology is broader than economics and politics. Many theoretical researches are difficult to be described by numerical value, but the development of sociology is gradually approaching numerical value.
    In the early period of sociology, Comte's positivism advocated the observation and classification of objects. Later, Marxism introduced the mathematical formula of economics into the study of sociology\cite{MarxKapital2011}, Émile Durkheim pioneered the introduction of statistics into sociology\cite{Durkheimsuicideetudesociologie1897}. Nowadays, big data has become one of the indispensable sociological research methods.
    The more suitable typical example here is Marxist theory. As can be seen from the picture, in the early society, the means of production was in line with the positioning of individuals. With the development of capitalism, the means of production moved towards companies, and even large projects reached the national level. This is an intuitive expression of Marxist theory.
\subsubsection{Philosophy and Religious Study}
    For religious philosophy, the author does not know much except some popular statements from Buddhism and Taoism.
    Referring to the philosophy of hermeneutics paradigms\cite{HusniHERMENEUTICSPARADIGMRELIGIOUS2018}, these statements can also be calibrated in TSQ-CS as a draft.
    For example, the Buddhist Scripture consider that there are tens of thousands lives in a bowl of water (Philosophy of Buddhism: sencence A, see on figure \ref{fig:map_all}). Taoism believes that life should be cherished and valued (Philosophy of Taoist sencence A). These philosophical thoughts can be calibrated
    to enrich these scriptures in TSQ-CS, and the scientific orientation of religious philosophy can be obtained with certain extent.
\subsubsection{Summary of Social Science}
    In a word, only by establishing a reliable mathematical model between groups and individuals can some sociological problems be solved optimally.
    For example, Matthew effect in economics, corruption in politics and administration, aggregation and division of countries and groups.
    Relying on TSQ-CS, scientists can identify some key needs like mathematical methods in sociology, and then focus on investment.

\section{Application of Science}
\hidsec{brief}
    According to the general distribution of current science, some general conclusions can be obtained, such as the worldview and methodology of science.
    The worldview can summarize the ability limit of current scientific research, and can also infer the maturity of various disciplines.
    Methodology includes more specific descriptions of various applications of formal science.
\subsection{Worldview of Science}
\hidsec{figure worldview}
    \begin{figure}[htbp]
    \centering
    \includegraphics[height=0.30\linewidth]{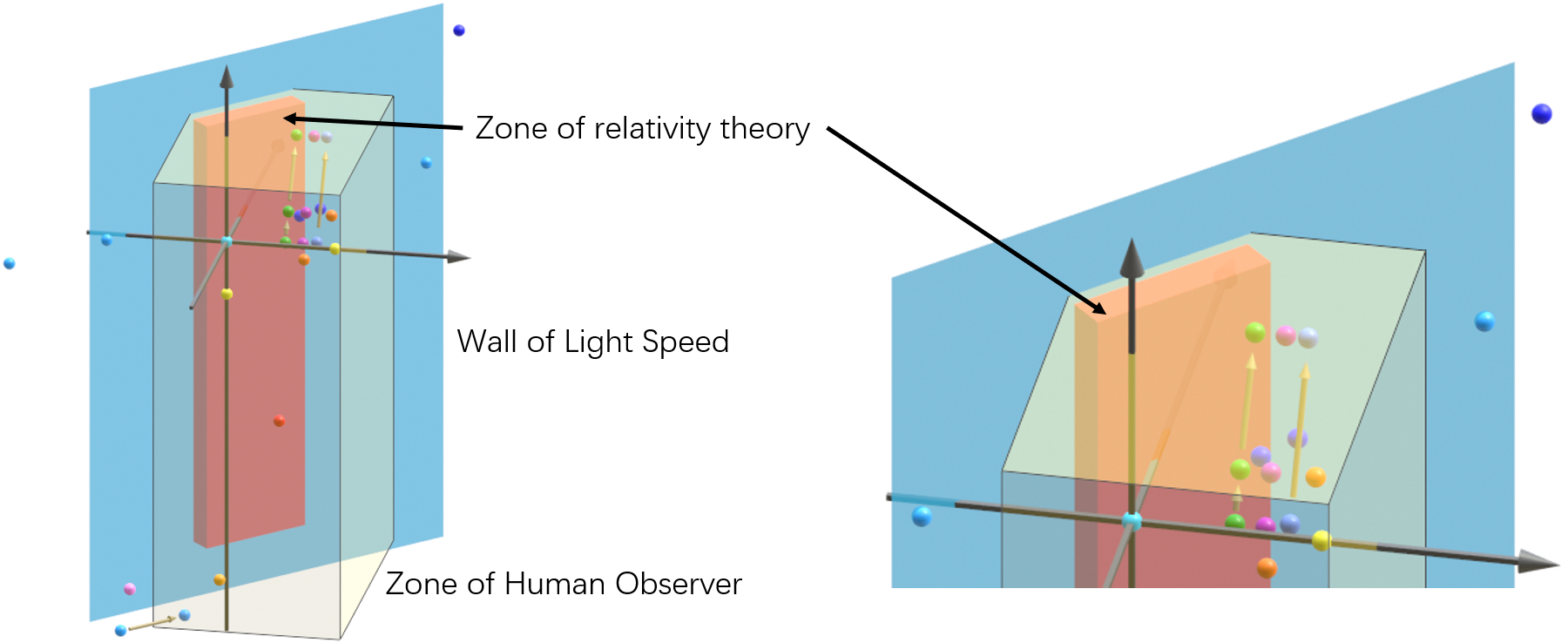}
    \caption{Worldview of Science}
    \label{fig:sci_definition}
    \end{figure}
\hidsec{undersdanding: current technology}
    In order to make the current science more clear, an observed zone $\Omega_{ob}$ can be defined in TSQ-CS as figure \ref{fig:sci_definition}, where
    \begin{equation}
        X \subset \Omega_{ob},\;\; X=\left[ \begin{array}{ccc} T & S & Q \end{array}\right]^{T}
    \end{equation}
    The range of $X_{ob}$ can not be provided accurately, but its components can be determined that
    \begin{equation} \label{eq:observe_zone}
        \begin{cases}
        N_{ob,max} \subset \left( N_{real-quantum}, N_{virtual-quantum} \right), & N =  T, S \; or \; Q \\
        N_{ob,min} \subset \left( N_{sun}, N_{galaxy} \right), & N = T, S \; or \; Q
        \end{cases}
    \end{equation}
    Eq \eqref{eq:observe_zone} can be used to represent the observed range of current science, in which the research objects are observed and used. The string theory, which is expected to unify physics, is still much far away.

    Taking light speed $c$ as the research object, and a wall of light speed can be obtained from a group of defined points. According to $s = ct$, let $\lambda _c = \lg c$, there is
    \begin{equation} \label{eq:speed_light_wall}
        X \subset \Omega_{wall}, \;\; where\;\; S = \lambda _c + T
    \end{equation}

    At present, all known disciplines are located on the same side of the wall of light speed, while the other side belongs to the unidentified superluminal range.
    For the scientific problems close to the plane, it needs to be corrected by the special theory of relativity.
\hidsec{target: need of maths}
    According to the figure \ref{fig:sci_definition}, the goal of scientific research can be simply summarized into three parts:
    \begin{enumerate}
    \item Discovery is to broaden the observed zone.
    \item Research is to acquire laws within observed zonee.
    \item Application is to apply acquired laws to human life.
    \end{enumerate}
\hidsec{difficulty}
    According to TSQ-CS, it can be predicted that the ultimate scientific theory of human expectation should not only be applicable to all scales of time and space, but also be able to adapt to the influence of number dimension.
\subsection{Methodology of Science}
\begin{figure}[htbp]
        \centering
        \begin{subfigure}{0.3\textwidth}
            \centering
            \includegraphics[height=.9\linewidth]{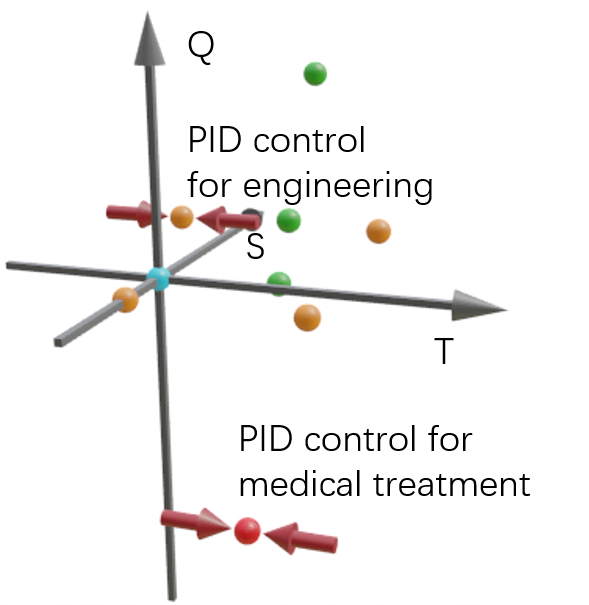}
            \caption{Approach in time}
            \label{fig:sub-1}
        \end{subfigure}
        \begin{subfigure}{0.3\textwidth}
            \centering
            \includegraphics[height=.9\linewidth]{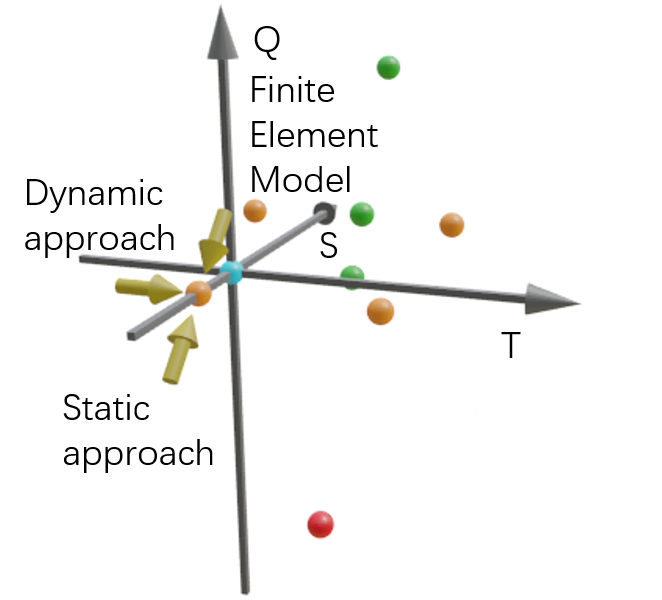}
            \caption{Approach in space}
            \label{fig:sub-2}
        \end{subfigure}
        \begin{subfigure}{0.3\textwidth}
            \centering
            \includegraphics[height=.9\linewidth]{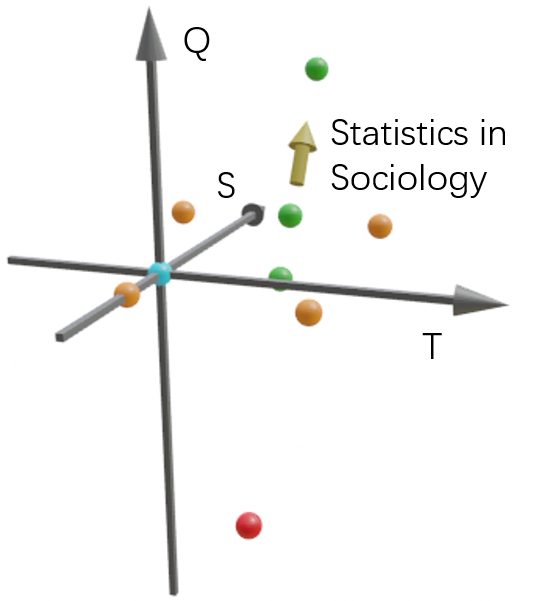}
            \caption{Approach in quantity}
            \label{fig:sub-3}
        \end{subfigure}
        \caption{Methodology of Science}
        \label{fig:application}
    \end{figure}

    Based on the categorization in last section, a general demonstration of scientific research can be concluded with the nodes and arrows. That is, to achieve a target state by moving on TSQ-CS with formal scientific methods from observable and measurable information.
    The approach can be kept as Eq \eqref{eq:general_approach}
    \begin{equation} \label{eq:general_approach}
        X_{target} = f_{method} \left( X_{1}, X_{2}, X_{3}, ... \right)
    \end{equation}
    Some samples of scientific method can be found as Fig \ref{fig:application}, similarity lies in methods belong to different disciplines. It should be noticed that there are 2 new points projected on plane $T=0$, which means the time durations are removed for solid body and assembled machinery.
\subsubsection{T-axis: PID Controlling in Time Dimension}
    PID is a classical control method in engineering, but its effective reason has not been exactly described.
    TSQ-CS can give the description of PID at the logic level.
    Assume $x = f\left( t, s, q\right)$, where $t = o(T)$, then $\text{d}x(t)/\text{d}t=o(T-1)$. Therefore, The PID components can be marked on TSQ-CS as Eq \eqref{eq:method_pid}, and described as Eq \eqref{eq:method_pid2}.
    \begin{equation} \label{eq:method_pid}
        \begin{cases}
            x(t) \rightarrow X(T) \\
            \text{d}x(t)/\text{d}t \rightarrow X(T - 1) \\
            \int x(t) \text{d}t \rightarrow X(T + 1)
        \end{cases}
    \end{equation}
    \begin{equation} \label{eq:method_pid2}
        X_{target}(T) = f_{PID} \left[ X_P(T), X_I(T+1), X_D(T-1) \right]
    \end{equation}
    According to the description, it is easy to see that the PID is a process of guiding or controlling a value with 2 values in nearby scales. This method enables the efficiency to obtain better results.

    In figure \ref{fig:application}, the controlling of medical treatment could be aided by PID same as engineering. Some applications of diabetes is already carried out\cite{CheeExpertPIDcontrol2003, MarchettiImprovedPIDSwitching2008}, and the PID method could be used in more medical area.
    Also, with TSQ-CS, PID can be applied to more disciplines similarly as above, such as body fit, country laws, etc.
    By the mean time, the analogy of various PID migration applications is helpful to develop its own theory.
\subsubsection{S-axis: Finite Element Anaylsis in Engineering}
    Finite Element Analysis(FEA) is one of the most mature engineering method, which can be a good demonstration how people surf on the TSQ-CS to approach the result they want.
    \begin{equation} \label{eq:fea}
        \begin{cases}
            X_{FEM}(T,S,Q) = f_{reform} \left( X_{geometry}(T, S+1, Q+1), X_{continuous}(T,S-1,Q-1) \right) \\
            X_{static}(T,S,Q) = f_{balance} \left( X_{FEM}, X_{boundary}(T,S+1,Q+1) \right) \\
            X_{dynamic}(T,S,Q) = f_{iteration} \left( X_{FEM}, X_{boundary}(T,S+1,Q+1), X_{initial}(T-1, S, Q) \right)
        \end{cases}
    \end{equation}
    Compared with the control of time scale in the world, calculus in the direction of S and Q axis is more used in analysis and calculation.
    As Fig \ref{fig:sub-2} and Eq \eqref{eq:fea}, the $f_{reform}$ is to discrete the geometry model, when the integral operation $f_{balance}$ in the S, Q direction is for static analysis, and the $f_{iteration}$ is for dynamic response simulation.
    Generally speaking, the integral from individual to overall level is a common operation in FEA.
    These three operations constitute a complete system, both clear in TSQ-CS and reliable in real industry.
\subsubsection{Q-axis: Distribution of Social Subjects}
    Similar science methods exist in the sociological research system which is completely different from industry.
    The study of quantitative scale change has existed for a long time in sociology.
    The methods is generally shown in Fig \ref{fig:sub-3}, and can be described as Eq \eqref{eq:soc_all}.
    \begin{equation} \label{eq:soc_all}
        X_{target}(T+n,S+n,Q+n)=f_{statistics} \left( X(T-n,S-n,Q-n) \right),\;\;
        \text{where:}\;n=\{0,1,2,...\}
    \end{equation}
    For example, classical positivist sociology\cite{HalfpennyPositivistsociologyits1994} focuses on the study of the whole and can be located at the highest sociology point in figure \ref{fig:sub-3}, and interpretative functionalist sociology, such as symbolic interactionism\cite{BlumerSymbolicinteractionismPerspective1986}, focuses on the study of individuals and locates at the lowest point.
    These ideas are different from each other, but finally constitute more comprehensive scientific view together.
    Such as modern structural functionalism\cite{KingsburyStructuralfunctionalism2009} and Marxist sociology\cite{MarxKapital2011} began to pay attention to the relationship between the whole and the individual, which could be coverd by the arrow from bottom to top.

    In fact, experts do have some research results in this field, which constructed a group of maths model. However, compared with the industrial methods, sociology is not only difficult to obtain a very reliable micro description, but also lack of unity for the overall external constraints and other macro level problems. These difficulties bring the flexiblity to Eq \eqref{eq:soc_all}, and kept it only in an high-level ideal form.
\subsubsection{Summary of Methodology} 
    Based on the discussion above, the methods of science in TSQ-CS can be summarised as follow:
    \begin{enumerate}
    \item The methods are in the observation zone, and they are more likely to be found and more mature near the human observer.
    \item Some methods like PID can be migrated to different subjects in different scales, and they both works.
    \item The methods along T and S axis represents better, when in Q axis, the social science is less comprehensible. Only a few methods like statistics are used, and more methods need to be developed for reversal operations.
    \end{enumerate}
    In a word, TSQ-CS is very suitable for describing various methods in scientific research intuitively, which can reflect the similarity between various methods in different disciplines.
    Therefore, a bridge can be built between various scientific methods, which can confirm and learn from each other.
    At the same time, it can also promote the comprehensive improvement of the research method and open up new application fields of the method.

\section{Conclusion}
\hidsec{purpose, method and result}
    This paper raised a new method for knowledge map in science, and obtained the TSQ-CS, which enables to locate disciplines and research targets with numerical properties.
    This method not only provides a worldview of science, but also helpful to conclude and migrating methodology between different subjects.
\hidsec{advantage}
    This TSQ-CS has the following advantages:
    \begin{enumerate}
    \item Categorization: according to the traditional way, the knowledge map is demarcated in the form of three labels, which makes the categorization more standardized and intuitive.
    \item Numerization: according to the modern way, the disciplines is demarcated in an objective form, which makes the knowledge map objective and unified instead of subjective and arbitrary.
    \item Arrow of formal science: bring the formal science into the numerical knowledge map, find their position in scientific research, and better serve the natural science.
    \item Quantity axis: it enables social sciences to be incorporated into knowledge map, interact with natural sciences, and provide opportunities for formal sciences to be applied.
    \item Worldview: a more comprehensive understanding of scientific system can be obtained from this, which is conducive to the understanding of science itself and the development of philosophy of science.
    \item Methodology: formal science can be more effectively applied in natural and Social Sciences, knowledge transfer can be generated between adjacent or similar disciplines.
    \item Foresight: current distribution of scientific knowledge could be found, so as to understand the current situation of human research. According to the vacancies and boundaries, future direction of research can be concluded, providing some references for the optimal allocation of research funds.
\end{enumerate}
\hidsec{limitation}
    There are also some limitations in TSQ-CS.
    \begin{enumerate}
    \item Number of tags: Although this framework can calibrate many disciplines, it can only be used at relatively high level. For specific scientific problems, there may be a lot of overlap on the coordinate axis, and can not be accurately described. In addition, the number of dimensions can not be more than three according to human spatial cognition.
    \item Numerical difficulty: this method adopts the basic principle of numerization, many disciplines are not applicable as they can not be numerized. For example, many traditional sub disciplines of sociology, the structure of graphene, the morphology of animals and plants and other issues, it is difficult to find labels that can be used for numerization.
    \end{enumerate}
\hidsec{summary}
    In a word, the purpose of this method is to provide some references for the application of philosophy of science and knowledge management in engineering and application view, so as to make the scientific research more comprehensive in vision, clearer in direction and closer in connection.

\section*{Acknowledgement}
\hidsec{acknoledgement}
    This research did not receive any specific grant from funding agencies in the public, commercial, or not-for-profit sectors.
    As the first draft of this paper was completed during career at Queen's University Belfast(QUB), the author is really grateful for the balance between job and life given by QUB, which allowed him think and write in relax after work. The subsequent revision was completed in the company of Wu Minghui, the author's girlfriend who provided love and spiritual support.
\hidsec{reference}
    \bibliographystyle{unsrt} 

\end{document}